\documentclass[journal,twocolumn]{IEEEtran}
\usepackage{amsfonts}
\usepackage{times}
\usepackage{graphicx}
\usepackage{latexsym}
\usepackage{dsfont}
\usepackage{amssymb}
\usepackage{amsmath}
\usepackage{cite}
\usepackage{verbatim}
\usepackage{subfigure}

\newcommand{\figref}[1]{{Fig.}~\ref{#1}}
\newcommand{\tabref}[1]{{Table}~\ref{#1}}

\def\bb0{{\mathbb{0}}}

\def\bb{{\mathbf{b}}}

\def\bff{{\mathbf{f}}}

\def\bh{{\mathbf{h}}}

\def\br{{\mathbf{r}}}

\def\bz{{\mathbf{z}}}
\def\b0{{\mathbf{0}}}

\def\bR{{\mathbf{R}}}

\def\bX{{\mathbf{X}}}

\def\sf0{{\mathsf{0}}}

\usepackage{epstopdf}
\usepackage{enumerate}
\usepackage{algorithm}
\usepackage{amsmath}
\usepackage{float}
\usepackage{color}
\usepackage{makeidx}
\usepackage{bm}
\usepackage{cleveref}
\usepackage{url}
\usepackage{steinmetz}
\usepackage{varwidth}
\usepackage{soul}
\graphicspath{{figs/}}
\usepackage{tabularx}
\usepackage{balance}

\newcommand{\sref}[1]{{Section}~\ref{#1}}

\DeclareMathOperator*{\argmin}{arg\,min}

\newcommand{\abs}[1]{\lvert#1\rvert}

\newcommand*{\thead}[1]{\multicolumn{1}{c|}{\begin{tabular}{@{}c@{}}#1\end{tabular}}}

\begin{document}

\title{Radar Aided Proactive Blockage Prediction in Real-World Millimeter Wave Systems}

\author{Umut Demirhan and  Ahmed Alkhateeb
\thanks{The authors are with the School of Electrical, Computer and Energy Engineering, Arizona State University, Tempe, AZ, 85281 USA (Email: udemirhan, alkhateeb@asu.edu). This work is supported by the National Science Foundation under Grant No. 2048021.}}

\maketitle

\begin{abstract}
	
	Millimeter wave (mmWave) and sub-terahertz communication systems rely mainly on line-of-sight (LOS) links between the transmitters and receivers. The sensitivity of these high-frequency LOS links to blockages, however, challenges the reliability and latency requirements of these communication networks. In this paper, we propose  to utilize radar sensors to provide sensing information about the surrounding environment and moving objects, and  leverage this information to proactively predict future link blockages before they happen. This is motivated by the low cost of the radar sensors, their ability to efficiently obtain important features such as the range, angle, velocity  of the moving scatterers (candidate blockages), and their capability to capture radar frames at relatively high speed. We formulate the radar-aided proactive blockage prediction problem and develop two solutions for this problem based on classical radar object tracking and deep neural networks. The two solutions are designed to leverage domain knowledge and the understanding of the blockage prediction problem. To accurately evaluate the proposed solutions, we build a large-scale real-world dataset, based on the DeepSense framework, gathering co-existing radar and mmWave communication measurements of more than $10$ thousand data points and various blockage objects (vehicles, bikes, humans, etc.).  	The evaluation results, based on this dataset, show that the proposed approaches can predict future blockages  $1$ second before they happen with more than  $90\%$ $F_1$ score (and more than $90\%$ accuracy). These results, among others, highlight a promising solution for blockage prediction and reliability enhancement in future wireless mmWave and terahertz communication systems.
\end{abstract}

\begin{IEEEkeywords}
	Radar, blockage prediction, machine learning, FMCW, mmWave, 6G
\end{IEEEkeywords}

\section{Introduction} \label{sec:introduction}
Future wireless networks attempt to meet the increasing demand on high data rates, low latency, and high reliability. More extensive usage of the higher frequency bands, millimeter-wave (mmWave) and sub-terahertz (sub-THz), is one prominent direction \cite{heath2016overview, rappaport2019wireless} for satisfying the high data rate demands.  However, the propagation characteristics at these frequencies result in two important features for mmWave/sub-THz communication systems: (i) These systems rely mainly on line-of-sight (LOS) links to guarantee sufficient receive signal power, and (ii) this dependency on LOS links coupled with the high penetration loss at mmWave/THz bands make these communication systems very sensitive to blockages. In particular, if these links are blocked, for example by moving objects, this could cause sudden performance degradation or even a link disconnection, which highly challenges the reliability and latency of these networks. This motivates the research for approaches that overcome the blockage challenges in high-frequency (mmWave/sub-THz) wireless networks. 

\begin{figure}[t]
	\centering
	\includegraphics[width=1\columnwidth]{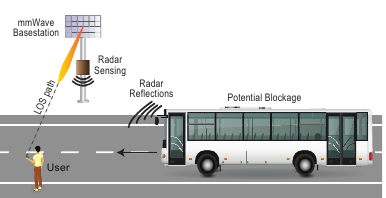}
	\caption{An illustration of the adopted system model. The mmWave LOS communication between the user and basestation is about to be interrupted by the moving bus due to the potential blockage of the LOS  path.  }
	\label{fig:systemmodel}
\end{figure}

Given that the LOS link blockages depend heavily on the positions of the communication terminals and the blockages in addition to their geometric properties (size, height, etc.), sensing the environment could potentially provide useful information for proactively predicting link blockages before they happen \cite{Alkhateeb_blockages}. Predicting future blockages enables the wireless network to make proactive decisions such as proactively handing off the user to another basestation or proactively switching to another beam.  In \cite{Alkhateeb_blockages}, beam sequences and recurrent neural networks were leveraged to predict future stationary blockages, and in \cite{alrabeiah2020deep}, the sub-6GHz channels were used to infer mmWave blockages. To enable dynamic (moving) blockage prediction, \cite{wu2021blockage} showed that the moving blockages produce  \textit{pre-blockage wireless signatures} that can potentially be leveraged for proactive blockage prediction. This approach, though, is mainly useful for scenarios with a few moving objects. To overcome these limitations and enable a more scalable solution, \cite{charan2021visionjournal} proposed to leverage red-green-blue (RGB) camera data which provide rich information about the moving objects in the surrounding environment. While cameras are relatively simple to deploy \cite{charan2021visionposition}, their usage is sometimes associated with privacy concerns and its operation may be limited scenarios with low light or bad weather conditions. Another promising approach is to use LiDAR sensory data, as proposed in \cite{wu2021blockageLiDAR}. Using LiDAR, however, is  mainly suitable for low-range scenarios and is associated with relatively high cost.

In this paper, we propose to leverage radar sensors to obtain useful information about the moving objects (candidate blockages) in the surrounding environment and use this information for proactive mmWave blockage prediction. The use of radar sensors (such as frequency-modulated continuous-wave (FMCW) radars) is motivated by (i) their off-the-shelf  availability at relatively low-cost, (ii) their capability to measure velocity (Doppler) in addition to range and angle, (iii) their potential high-frequency low-latency measurements and (iv) the lack of privacy concerns with radar sensory data, and (v) the possibility to be integrated with communication systems \cite{Taha2021,Kumari_2018}. With this motivation,  we formalize the radar-aided proactive blockage prediction problem and present two alternative solutions: One is based on classical signal processing for radar object tracking and one that leverages deep neural networks. In the first approach, we develop a radar object tracking algorithm based on Unscented Kalman filter \cite{wan2000unscented} whose states are later utilized with a k-nearest neighbors classification algorithm for blockage prediction. In the second approach, we utilize the range-angle maps obtained from the radar measurements and develop a  deep neural network based solution to predict future blockages. 

To evaluate the performance of the proposed solutions, we build a large-scale real-world dataset for an outdoor scenario, based on the DeepSense framework \cite{DeepSense}, with coexisting radar and mmWave communication data. The constructed dataset comprises around $10$ thousand data points from more than  $300$ unique blockage trajectories including vehicles, bikes, and humans. With this dataset, our evaluation results show that the deep learning approach gives promising blockage prediction accuracy gains with the advantage of design simplicity. For example, the results indicate that the developed solutions could  predicted future blockages  $1$ second before they happen with more than  $90\%$ $F_1$ score (and more than $90\%$ accuracy). These results, among others, highlight a promising solution for blockage prediction and reliability enhancement in future wireless communication systems.

The rest of the paper is organized as follows. In \sref{sec:systemmodel}, the system model is presented. Building upon the system model, \sref{sec:problemformulation} formulates the radar-aided blockage prediction problem. \sref{sec:trackingsol} and \sref{sec:solution-ml} propose the object tracking and the deep learning solutions, respectively, for the radar-aided blockage prediction problem. For the evaluation, \sref{sec:dataset} details the experimental setup and real-world dataset, and \sref{sec:results} presents the numerical results utilizing this dataset. Finally, the paper is concluded in \sref{sec:conclusion}.

\section{System Model} \label{sec:systemmodel}

The considered system comprises a basestation communicating with a stationary user. The basestation adopts two main components: (i) A mmWave communication transceiver equipped with a phased array communicating with the stationary user and (ii) an FMCW radar. The radar is utilized to sense the environment and predict the LOS link blockages that are caused by the objects (e.g., vehicles) moving between the base station and user. An illustration of the system model is shown in \figref{fig:systemmodel}. In the following two subsections, we briefly describe the system and signal models of the adopted communication and radar components.

\subsection{Radar Model}
In our system, the basestation is equipped with an FMCW radar. The radar device provides measurements for the communication environment around the base station, which could be leveraged for predicting future blockages. The radar captures one measurement every $\tau_f$ seconds. In each measurement, the FMCW radar transmits a frame of $L$ chirps. Each chirp has a linearly increasing frequency starting at an initial frequency $f_c$ and ending at a stop frequency $f_c + \mu t$, given by
\begin{equation}
s^\textrm{tx}_\textrm{chirp}(t) = 
\begin{cases}
\sin( 2\pi [f_c \, t + \frac{\mu}{2} \, t^2]) & \text{if }  0\leq t \leq \tau_c, \\
0 & \text{otherwise},
\end{cases}
\end{equation}
where $\mu= B/\tau_c$ is the slope of the linear chirp signal with $B$ and $\tau_c$ representing the bandwidth and duration of the chirp, respectively. As mentioned, each radar measurement is collected from a frame of $L$ chirps, and the chirps are transmitted with a waiting time $\tau_s$ between them. After the transmission of $L$ chirps, no other signals are transmitted until the next frame. The transmit signal of the radar frame can be written as
\begin{equation}
s^\textrm{tx}_\textrm{frame}(t) = \sqrt{\mathcal{E}_t} \sum_{l=0}^{L-1} s_\textrm{chirp}(t - (\tau_c+\tau_s)\cdot l), \quad 0\leq t\leq \tau_f
\end{equation}
where $\sqrt{\mathcal{E}_t}$ is the transmitter gain. 

The radar transmit signal is reflected on the different objects in the environment, and is received back at the radar. At the receiver, the signal obtained from an antenna is passed through a quadrature mixer that combines the transmit with receive signals, producing the in-phase and quadrature components. After that, a low-pass filter is applied to the mixed signals. The resulting signal, referred to as intermediate frequency (IF) signal, reflects the frequency and phase difference between the transmit and receive signals. If a single object exists in the environment, then the receive IF signal of a single chirp can be written as \cite{iovescu2017radarfundamentals}
\begin{equation}
s_\textrm{chirp}^\textrm{rx}(t) = \sqrt{\mathcal{E}_t \mathcal{E}_r} \exp\left(j2\pi \left[\mu \tau_{rt} t + f_c \tau_{rt} - \frac{\mu}{2} \tau_{rt}^2\right]\right),
\end{equation}
where $\sqrt{\mathcal{E}_r}$ is the channel gain of the object which depends on the radar cross section (RCS) and path-loss, $\tau_{rt} = 2d/c$ is the round-trip delay of the signal reflected from the object. The symbol $d$ denotes the distance between the object and the radar, and $c$ represents the speed of light. 

The receive IF signal, $s_\textrm{chirp}^\textrm{rx}(t)$ is then sampled at the sampling rate of the ADC, $f_s$, producing $S$ samples for each chirp. Given the $L$ chirps per frame, and assuming an FMCW radar with $M_r$ receive antennas (with an RF chain per antennas), each radar measurement produces $M_r \cdot S \cdot L$ ADC samples. We use $\bR \in \mathbb{C}^{M_r\times S\times L}$ to denote the receive radar ADC samples (raw data) of each measurement. Please refer to \cite{iovescu2017radarfundamentals, dham2017programming} for more information about the adopted FMCW radar and its hardware architecture. Next, we describe the communication and blockage models.

\subsection{Communication and Blockage Models}
The considered base station employs a mmWave transceiver with $M_\mathrm{A}$ antennas and use it to communicate with a single-antenna mobile user. We adopt a narrowband channel model and write the channel between the base station and user as
\begin{equation}\label{eqn:channel}
\widetilde{\bh} = \widetilde \bh^\mathrm{LOS} + \widetilde\bh^\mathrm{NLOS}, 
\end{equation}
where $\bh^\mathrm{LOS}$ and $\bh^\mathrm{NLOS}$ are the channel coefficients due to the LOS and NLOS paths. At the downlink, the basestation utilizes the beamforming vector $\bff \in \mathbb{C}^{M_A}$ to transmit the symbol $s_d$ to the user. With this model, the receive signal at the user can be expressed as
\begin{equation}
	y = \sqrt{\mathcal{E}_c} \, \widetilde\bh^H \bff s_d + n, 
\end{equation}
where $\sqrt{\mathcal{E}_c}$ is the transmit gain of the basestation, and $n \sim \mathcal{CN}(0, \sigma^2)$ is the additive white Gaussian noise with $\sigma^2$ being the variance. The beamforming vector $\bff$ is assumed to be selected from a pre-defined codebook $\boldsymbol{\mathcal{F}}$, i.e., $\bff \in \boldsymbol{\mathcal{F}}$ \cite{Zhang2021}. In particular, the basestation selects the optimal beamforming vector $\bff^\star$ that maximizes the receive beamforming gain $|\bh^H \bff|^2$. In this work, Assuming that $f^\star$ is selected, we write the effective channel as
\begin{equation}
	h = \widetilde\bh^H \bff^\star = h^\mathrm{LOS} + h^\mathrm{NLOS}, 
\end{equation}
with $h^\mathrm{LOS}$ and $h^\mathrm{NLOS}$ are the effective channel gains of the LOS and NLOS components.

\textbf{Incorporating Blockages:} Adopting a block fading channel model,  we define $h[t]$ and $\bR[t]$ as the channel gain and radar measurements at time instance $t \in \mathbb{Z}^+$. Now, we can define the blockage indicator at time instance $t$ by $\widetilde{b}[t] \in \{0, 1\}$, which indicates the LOS path being blocked ($\widetilde b[t]=1$) or not ($\widetilde b[t]=0$). With the blockage indicator, we can write the channel gain at time instance $t$ as
\begin{equation}
	h[t] = (1-\widetilde{b}[t]) \cdot h^\mathrm{LOS}[t] + h^\mathrm{NLOS}[t].
\end{equation}
We note that for the mmWave and THz frequency communication bands considered in this paper, there are usually a limited number of NLOS paths \cite{rappaport2013millimeter}, and the channel gains with the blockage are comparably smaller, i.e., $\abs{h^\mathrm{LOS}} \gg \abs{h^\mathrm{NLOS}}$. Next, we formulate the blockage prediction problem.

\section{Radar Aided Blockage Prediction:\\ Problem Formulation} \label{sec:problemformulation}
In this section, building upon the system model described in \sref{sec:systemmodel}, we  define the radar-based blockage prediction problem. This paper aims to predict future blockages utilizing the current and previous radar measurements. Formally, we consider the latest (past and present) $T_o$ radar observations to predict a blockage within the next $T_p$ time-slots. Let us denote the set of the $T_o$ latest radar measurements by
\begin{equation}
\bX[t] = \left\{\bR[t-T_o+1],\ \ldots,\ \bR[t] \right\}.
\end{equation} 
With this information, our purpose is to predict the blockage status in the following $T_p$ time-slots, i.e., $\{t+1, \ldots, t+T_p\}$. If there is any blockage during these slots, the blockage status for the $T_p$ slots, $b[t]$, is considered as blocked. Mathematically, we can write
\begin{equation}
	b_t = \bigvee_{t_p=1}^{T_p} \widetilde b[t+t_p], 
\end{equation}
where $\vee$ is the logical \textit{OR} operation. With this notation, we define a function $\Psi_{\bf \Theta}$ that maps the stack of the radar measurements $\bX[t]$ to the blockage status $b[t]$. Mathematically, we can write
\begin{equation} \label{eqn:psidef}
	\Psi_{\bf \Theta}: \bX[t] \rightarrow b[t].
\end{equation}
The function $\Psi_{\bf\Theta}$ (with its parameters $\bf\Theta$) returns the blockage status given the radar measurements. Hence, our purpose in this paper is to design a function $\Psi_{\bf\Theta}$ that approximates the function defined in \eqref{eqn:psidef} and the optimization of its parameters, $\bf \Theta$. With the optimal function and parameters being denoted by $\Psi^\star$ and $\bf \Theta^\star$, we can formalize the problem by
\begin{equation} \label{eqn:problemdef}
	\Psi_{\bf{\Theta}^\star}^\star = \argmin_{\Psi_{\bf\Theta}} \frac{1}{T} \sum_{t=1}^T \mathcal{L}(\Psi_{\bf\Theta}(\bX[t]), b[t]), 
\end{equation}
where $T$ is the total number of time-slot samples and $\mathcal{L}(.,.)$ is the loss function of the predictions. In the next two sections, we present two proposed solutions for radar-aided blockage prediction: One is based on radar object tracking and the other one is based on deep learning.

\section{Radar Aided Blockage Prediction:\\An Object Tracking Solution} \label{sec:trackingsol}

In this section, we propose a radar-aided blockage prediction approach based on target/object tracking. The proposed solution applies the following processing pipeline on the radar measurements: (i) Obtaining range, velocity, and angle Fast Fourier Transformations (FFTs), (ii) detecting the targets by applying a constant rate false alarm rate (CFAR) algorithm \cite{rohling1983radar}, (iii) applying a density-based spatial clustering of applications with noise (DBSCAN) algorithm  \cite{ester1996density, schubert2017dbscan} for clustering the detected points, (iv) determining the  range/angle/velocity measurements of the detected/clustered objects, (v)  associating the measurements of the objects detected in different time samples to form tracks,  (vi) tracking the object states (position/velocity) with a Kalman filter, and (vii) predicting the future blockages based on the states (predicted position/velocity) of the objects. Next, we detail the various steps of the proposed approach.

\textbf{Pre-processing:} The radar measurements $\bR[t]$ are first processed to obtain range, angle and velocity information. In this method, we apply the pre-processing to initially obtain the radar cube of the range, angle and velocity which is used for the detection of the moving blockage object. To extract the range, angle and velocity information from a radar measurement $\bR[t]$, three FFTs are applied. In summary, (i) an FFT in the direction of the time samples, referred to as the range FFT, is applied to obtain the range information, (ii) an FFT through the chirp samples, called as the Doppler FFT, is applied for the velocity information, and (iii) for the angle information, an FFT through the direction of the antenna samples, referred to as the angle FFT, is applied. We consider the FFTs of size $N_S$, $N_L$, and $N_M$, and denote the resulting processed radar information by $\bR_\mathrm{RC}[t] \in \mathbb{C}^{N_M \times N_S \times N_L }$, referred to as the radar cube. If we denote the 3D FFT by $\mathcal{F}_\mathrm{3D}$, the radar cube can be mathematically expressed as $\bR_\mathrm{RC}[t]=\mathcal{F}_\mathrm{3D}(\bR[t])$. In the radar cube, the 2D matrices for each angular sample, transformed from the antenna samples by the angle FFT, contain the range-velocity maps. This information can be further reduced by summing over different angular samples. To clarify, we can write the transformation for the range-velocity maps, denoted by $\bR_\mathrm{RV} \in \mathbb{R}^{N_M \times N_S}$, as
\begin{equation}
	\bR_\mathrm{RV}[t] = \sum_{n=1}^{N_M}\abs{ \bR_\mathrm{RC}[t]_{(n, :, :)}) }, 
\end{equation}
where the sub-indices of the radar cube denotes the corresponding elements of the cube. Now, the resulting range-velocity maps are ready for the detection of the objects.

\textbf{Detection:} For the detection of the points in the radar measurements, the constant rate false alarm rate (CFAR) algorithm \cite{rohling1983radar} is applied. In the CFAR algorithm, a window and guard region is determined. For the cell-averaging, the average of the points within the window around the cell under test, excluding the guard-cells, are averaged. This average is multiplied by a constant to construct the detection threshold for the cell under test.
In our approach, we adopt a comparably simple 2D cell-averaging CFAR on the range-velocity map to detect the points with reflections. We select the constant for the threshold by inspection. Then, we apply the algorithm for each point in the range-velocity map, and collect the detected points in a binary range-velocity map.

\textbf{Clustering:} After applying the CFAR detection algorithm, the detected points for the different objects need to be determined. To detail, the detected point do not necessarily present a single point for each single object. In addition, due to the noise and clutter, the points are not necessarily completely grouped around the objects. Hence, the detected points of the map need to be grouped by a clustering algorithm with unknown number of clusters (objects). For this purpose, we apply the DBSCAN \cite{ester1996density, schubert2017dbscan} clustering algorithm to group the detected points into candidate link blockage objects.

\textbf{Measurements of the objects:} For each object returned by the DBSCAN algorithm, a measurement vector needs to be obtained for tracking. As the measurements, range, velocity and angle information can be extracted from the available radar cube. For the range and velocity, each detected point of an object in the binary range-velocity map correspond to a range and velocity pair. As the range and velocity of this object, we take the average of these pairs. For the angle of an object, first, the angle slices of the 3D radar cube, that correspond to the detection points of the object, are summed. Then, in this sum, the angle corresponding to the bin with the maximum value is taken as the angle measurement of the object.

\textbf{Data association:} With the range, angle and velocity measurements of each object extracted from the radar measurements, the objects in consecutive time samples are associated to construct the tracks of the objects. In particular, we resort to a weighted metric as the distance, which is the weighted sum of the Euclidean position (obtained by using the range and angle), and velocity. Based on this metric, the closest objects in the consecutive time samples are associated with each other, unless the distance between the objects does not exceed a pre-determined threshold value. To clarify, the measurements of the object may not be available in the following frame, and it should not be associated with any other detected object based on the distance. The threshold aims to ensure this is the case. Furthermore, if a track of data is not associated with any object in the last $t_e$ measurements, it is assumed to be ended and removed from the list of active tracks. Next, we apply a Kalman filter to the constructed object tracks.

\textbf{Kalman filter:}  In this step, the noisy measurements of the tracks are passed through a Kalman filter to estimate the states of the objects. Specifically, the Kalman filter adopts the state transition and observation/measurement models to estimate the current state by utilizing the previous state and the current measurements. To detail, the state update model is utilized for the prediction of the current state from the previous state. Then, via the measurement model, the measurements corresponding to the predicted next state are estimated. After obtaining the actual measurement, the difference between the prediction and actual value is utilized for the update step of the Kalman filter, providing a robust estimate for the current state.

For the state transition model of this filter, we follow the constant velocity model \cite{schubert2008comparison}, and define the states of the objects as the position and velocity in the Cartesian coordinates with the notation $\bz[t]= [x[t], y[t], v_x[t],v_y[t]]$. In the constant velocity model, the state is updated as 
\begin{equation} \label{eqn:statemodel}
	\bz[t+1] = \big[x[t]+\tau v_x[t],\ v_x[t],\ y[t]+\tau v_y[t],\ v_y[t]\big],
\end{equation}
For the measurement model, we adopt the augmented measurement model proposed in \cite{vaishnav2020continuous}. In this model, the range, angle and velocity are the measurements, and the Cartesian position and velocity are the states. Mathematically, it is written by
\begin{align}
	\rho = \sqrt{x^2+y^2}, \\
	v = \frac{x v_x + y v_y}{\sqrt{x^2+y^2}}, \\
	\theta = \arctan{\frac{y}{x}},	
\end{align}
where $\rho, v$ and $\theta$ denote the range, velocity and angle measurements, respectively. 

In this work, due to the non-linearity of the observation model, an Unscented Kalman filter \cite{wan2000unscented} is adopted for the tracking of the objects. For the multiple active objects, different Kalman filters initiated. To distinguish, we denote the number of objects tracked with separate Kalman filters at time instance $t$ by $O[t]$, and the estimated state of the object $o \in  \{1, \ldots, O[t]\}$ by $\bz^o[t]$.

\textbf{Blockage Prediction:} For the prediction of the blockages, different types of methods can be considered. For instance, in a more interpretable manner, the current states of the objects can be projected to the future blockage instances via the constant velocity model of the Kalman filter \eqref{eqn:statemodel}, and a region to determine the blockage status for each of these future instances can be selected. Nevertheless, from a data-centric and potentially robust manner, we aim to classify the blockage status based on the current object states. Specifically, the states of the objects are stacked with the addition of zeros padding in a vector given by 
\begin{equation}
	\bX_{track}[t] = \big[\bz^1[t], \ldots, \bz^{O[t]}[t], \b0, \ldots, \b0 \big], 
\end{equation}
where $\bX_{track}[t]\in \mathbb{R}^{4\cdot\max_t O[t]}$. With this definition, we can finally apply the classification method that aims to return $b[t]$ from the given states $\bX_{track}[t]$. For the classification, we empirically test several classical machine learning classifiers, and adopt k-nearest neighbor (k-NN) algorithm that performs the best over the dataset.

\section{Radar Aided Blockage Prediction:\\ A Deep Learning Solution} \label{sec:solution-ml}

\begin{figure*}[ht]
	\centering
	\includegraphics[width=1.9\columnwidth]{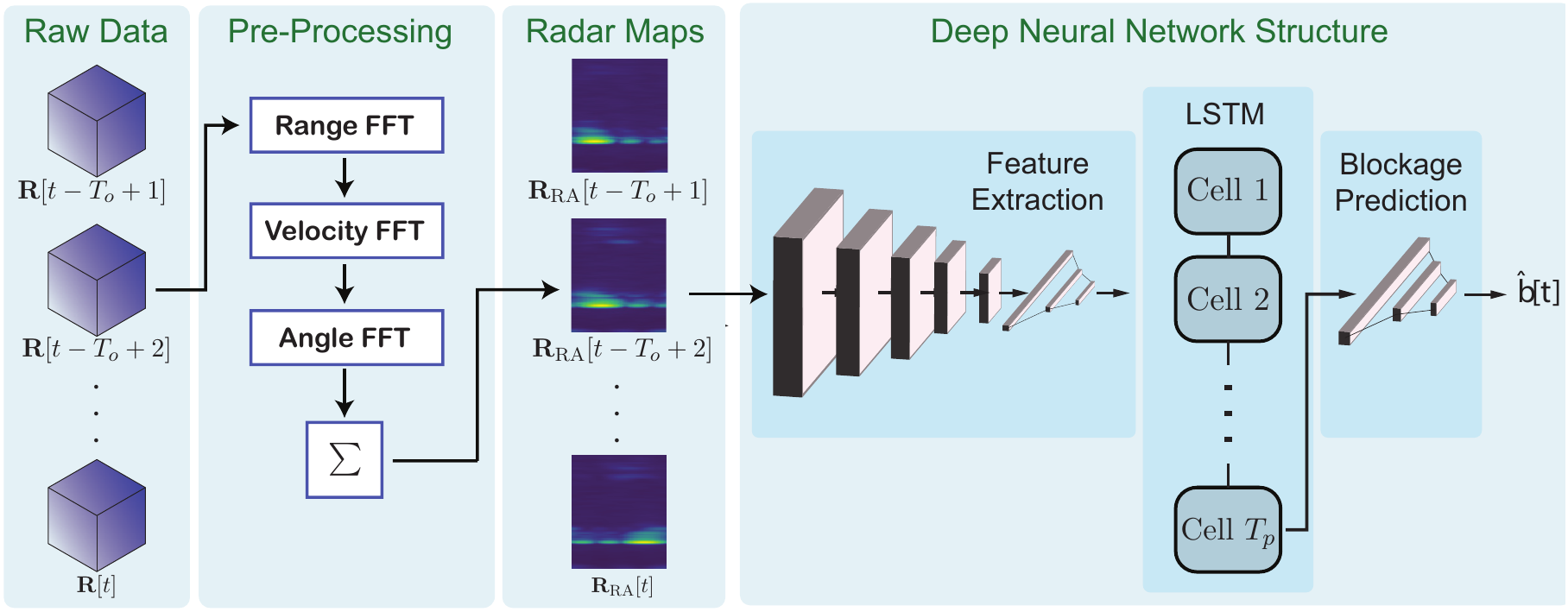}
	\caption{The schema of the proposed deep learning solution. The raw data is first processed to obtain the range-angle maps by applying the signal processing techniques. The series of the range-angle maps are then jointly fed to a neural network that comprises feature extraction, LSTM and blockage prediction sub-networks.}
	\label{fig:ml_solution_model}
\end{figure*}

In this section, we propose a deep learning solution for the radar-aided blockage prediction problem defined in \sref{sec:problemformulation}. The solution learns a mapping function  $\Psi$, in \eqref{eqn:problemdef}, and its variable parameters to solve the considered problem. Our solution in this section does not just adopt a deep neural network model, but also incorporate a domain-knowledge pre-processing approach to present interpretable inputs to the learning model. Specifically,  the considered approach starts by pre-processing the radar measurements to extract the range-angle maps. Then, leveraging the understanding of the object tracking problem, we  construct a deep neural network architecture, consisting of three stages that aim to extract the relevant features, exploit the sequential correlation, and make classification decisions. This is done via a combination of convolutional neural networks (CNNs), LSTM networks, and fully-connected layers, which complement each other and present a promising solution. In particular, the range-angle maps are fed to the developed model that comprises (i) a CNN-based feature extraction component to extract the essential information from the maps, (ii) an LSTM component to take advantage of the time correlations, and (iii) a linear prediction layer to return the blockage status. The proposed solution, along with its components, are depicted in \figref{fig:ml_solution_model}. Next, we  detail the components of the developed solution, namely, pre-processing; feature extraction, LSTM, and prediction layers.

\textbf{Pre-processing:} In this step, we follow a similar pre-processing approach to that proposed in \sref{sec:trackingsol}. In summary, we apply the range, velocity and angle FFTs of size $N_S$, $N_L$, and $N_M$, respectively, and obtain the radar cube $\bR_\mathrm{RC}[t] \in \mathbb{C}^{N_M \times N_S \times N_L }$ from the measurement $\bR[t]$. In the radar cube, the 2D matrices for each chirp sample contains the range-angle maps, which can be further reduced by summing over different chirp samples. Thus, we can write the transformation for the range-angle maps, denoted by $\bR_\mathrm{RA} \in \mathbb{R}^{N_M \times N_S}$, as
\begin{equation}
\bR_\mathrm{RA}[t] = \sum_{n=1}^{N_L} \abs{ \bR_\mathrm{RC}[t]_{(:, :, n)}) }. 
\end{equation}
\noindent The resulting range-angle maps of different time samples are ready for the deep learning processing.

\textbf{Feature Extraction:} The range-angle maps contain information about the receive power levels for each point in the maps. However, not all of this information is useful for the prediction of the blockages since only the moving objects are relevant and may cause a blockage. Hence, the irrelevant information can be minimized by extracting  a smaller number of features from the range-angle maps. This lower-dimensional representation helps to ease the complexity of the following LSTM architecture. To exemplify a similar approach from the video processing, an object detector network can be adopted to find the objects before tracking them through the LSTM layers. In addition, this dimensionality reduction does not necessarily degrade the performance \cite{sainath2015convolutional}.

Therefore, as the first part of the neural network, we adopt a CNN architecture to reduce the dimensionality and extract the essential local features, relying on the powerful capabilities of CNN networks in similar complex tasks \cite{albawi2017understanding}. For this network architecture, we adopt a sequence of the convolutional, average pooling and fully-connected layers. In our design, this part of the network does not aim to extract time-dependent information, and hence, the range-angle maps from different time samples can be fed to the same network separately. Similarly, the network can be trained with the gradient due to the each output separately, providing more samples and faster training opportunity. Formally, the network takes a single range-angle map $\bR_\mathrm{RA}[t]$ as the input, passes the input through its layers, and returns the extracted features of this map, $\br_\mathrm{RA}[t]$, as the output. The architecture of the feature extraction network is summarized in \tabref{table:featureextractionnet}, which is adopted from the deep learning model that has shown successful results \cite{demirhan2021radarbeam} for the range-angle map beam prediction task. 

\begin{table}[!t]
	\centering
	\caption{Architecture of the Feature Extraction Network }
	\label{table:featureextractionnet}
	\renewcommand{\arraystretch}{1.3}
	\resizebox{\columnwidth}{!}{
		\begin{tabular}{|c|c|}
			\hline
			\textbf{NN Layers} & \textbf{Range-Angle ($\bR_{\mathrm{RA}}$)}        \\ \hline
			\textbf{Input}      & $1 \times 256 \times 64$     \\ \hline
			\textbf{CNN-1}     & Output Channels: 4, Kernel: (3, 3), Activation: ReLU  \\ 
			\hline
			\textbf{CNN-2}     & Output Channels: 8, Kernel: (3, 3), Activation: ReLU  \\
			\hline
			\textbf{CNN-3}     & Output Channels: 16, Kernel: (3, 3), Activation: ReLU  \\ 
			\hline
			\textbf{AvgPool-1} & Kernel: (2, 2)       \\ \hline
			\textbf{CNN-4}     & Output Channels: 4, Kernel: (3, 3), Activation: ReLU  \\ 
			\hline
			\textbf{AvgPool-2} & Kernel: (2, 2)       \\ \hline
			\textbf{CNN-5}     & {Output Channels: 2, Kernel: (3, 3), Activation: ReLU}  \\ 
			\hline
			\textbf{AvgPool-3} & {Kernel: (2, 2)}       \\ \hline
			\textbf{FC-1}     & {Input Size: 512, Output Size: 256, Activation: ReLU}  \\ 
			\hline
			\textbf{FC-2}     & {Input Size: 256, Output Size: 64, Activation: ReLU}  \\ 
			\hline
		\end{tabular}	
	}
\end{table}

\textbf{Long-Term Short-Memory Networks:} After the feature extraction, the dependency of the features across  the different time samples can be captured. For this purpose, the family of recurrent neural networks, which contains sequential connections between the cells of different inputs, can be utilized. In this work, we adopt the LSTM \cite{gers2000learning} networks due to their successful radar applications such as object tracking and classification \cite{akita2019object} and hand gesture recognition \cite{choi2019short}. 

To detail, an LSTM network consists of multiple LSTM cells, each taking a single entry of the time-sequence data. These LSTM cells are connected to each other in a sequential manner, and each can return an output vector, resulting in an output sequence of these vectors. For the blockage prediction, we only adopt the vector returned from the latest cell. Formally, the network takes  $\{\br_{RA}[t-v+1], \ldots, \br_{RA}[t]\}$ as the input, and the last cell returns the intermediate output, $\widetilde{\mathbf{\br}}[t]$ of the size of $\br_{\mathrm{RA}}[t]$, to be utilized by the classification network.

\textbf{Blockage Prediction:} Finally, the output of the LSTM is fed to another set of fully-connected neural network layers to obtain the prediction of the blockage. The input of the blockage prediction layers is of the size of a feature vector extracted from a single range-angle map. For the prediction of the blockage, a final set of layers returning the blockage prediction output is required. For this purpose, we utilize a simple neural network of a single fully-connected layer. 
This final layer only returns a single prediction as a soft information, i.e., ${\hat{b}'}[t] \in [0, 1]$, which is later converted to the binary value of the blockage prediction, $\hat{b}[t]=\mathbf{1}\{{\hat{b}'}[t]>0.5\}$.

\textbf{Neural Network Training and Loss Function:}
The neural network is trained by adopting the formulation of \eqref{eqn:problemdef}, only over the parameters $\bf\Theta$. To clarify, as the design of the function $\Psi$ is fixed by the proposed process, we only aim to find the parameters that minimize the loss over the data samples by
\begin{equation} \label{eqn:nnobjective}
	{\bf \Theta^\star} = \argmin_{\bf \Theta} \frac{1}{L} \sum_{t=1}^T \mathcal{L}\left(\hat{b}'[t], b[t]\right),
\end{equation}
where the loss function is defined as the binary cross-entropy as the problem is a binary classification problem. The binary cross-entropy function can be written by
$$\mathcal{L}(b[t], \hat{b}'[t]) = b[t]\log(\hat{b}'[t]) + (1-b[t])\log(1-\hat{b}'[t]).$$
For the neural network objective defined in \eqref{eqn:nnobjective}, the error can be computed at the output of the network, and it can be back-propagated through the layers with gradient descent (or alternative) methods, optimizing the parameters $\bf\Theta$.

\section{Experimental Setup and Real-World Dataset} \label{sec:dataset}

\begin{figure*}[!t]
	\centering
	\subfigure[The basestation antenna array and radar device on the setup]{
		\includegraphics[height=.27\linewidth]{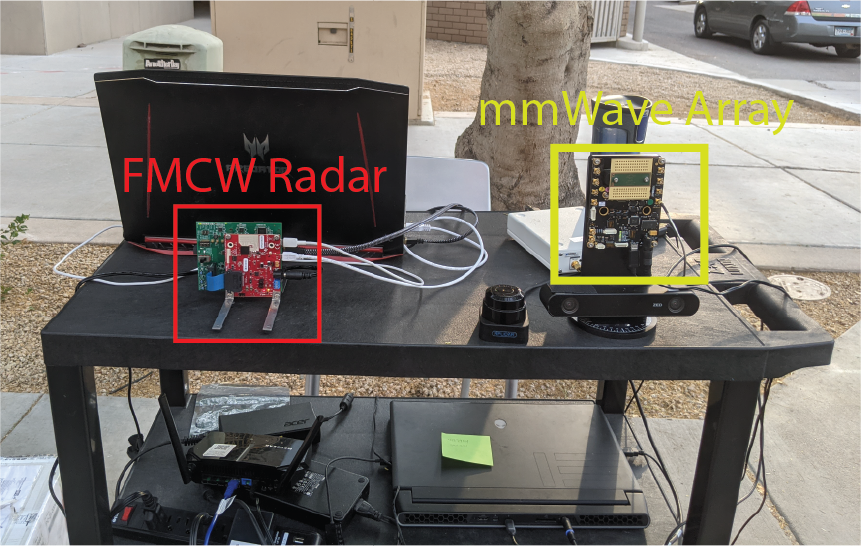}
		\label{fig:setup_im_a}
	}
	\subfigure[The system setup with a car on sight]{
		\includegraphics[height=.27\linewidth]{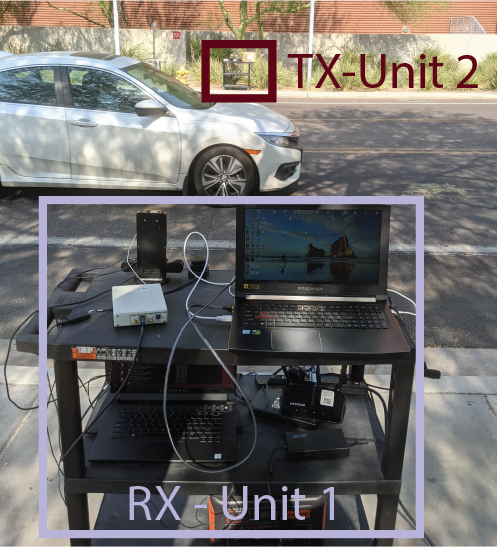}
		\label{fig:setup_im_b}
	}
	\subfigure[Satellite image of the scenario]{
		\includegraphics[height=.27\linewidth]{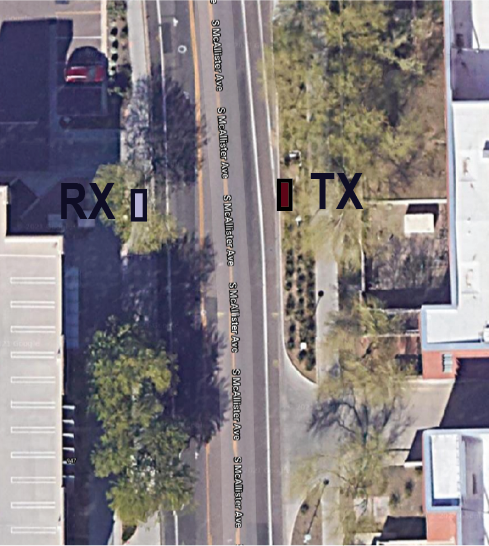}
		\label{fig:setup_im_c}
	}
	\caption{The testbed and scenario details are shown from (a) the front-view, (b) the back-view with a potential blockage car on sight, (c) the satellite. In (a), a closer look on the testbed with mmWave array and FMCW radar is provided. As shown in (b) and (c), the transmitter is located on the east side of the road while the receiver on the west. 
	}
	\label{fig:datasetup}
\end{figure*}

\begin{table}[t]
	\caption{Scenario 30: McAlister Stationary Blockage}
	\label{table}
	\centering
	\setlength{\tabcolsep}{3pt}
	\renewcommand{\arraystretch}{1.4}
	\begin{tabular}{|c|c|}
		\hline\hline
		\textbf{Testbed}             & 3               \\ \hline
		\textbf{Number of Instances}  & 76000               \\ \hline                     \hline
		\textbf{Number of Units}     & 2 \\ \hline
		\textbf{Total Data Modalities}     & \thead{RGB images, LiDAR and radar \\ 64-dimensional received power vector \\ GPS locations} \\ \hline \hline
		\multicolumn{2}{|c|}{\textbf{Unit 1}} \\ \hline
		\textbf{Type} & Stationary \\ \hline
		\textbf{Hardware elements} & \thead{RGB camera, LiDAR, radar \\ mmWave phased array receiver \\ GPS receiver} \\ \hline
		\textbf{Data Modalities} & \thead{RGB images, LiDAR and radar measurements \\ 64-dimensional received power vector \\ GPS location }  \\
		\hline \hline
		\multicolumn{2}{|c|}{\textbf{Unit 2}} \\ \hline
		\textbf{Type} & Stationary \\ \hline
		\textbf{Hardware elements} & \thead{mmWave omni-directional transmitter, \\ GPS receiver} \\ \hline
		\textbf{Data Modalities} & GPS location \\
		\hline\hline
	\end{tabular}
	\label{table:scenario30}
\end{table}

For a realistic evaluation of the proposed radar and machine learning aided blockage prediction solution, we collected a large-scale real-world  dataset using a hardware testbed with co-existing radar and wireless mmWave equipment, following the DeepSense dataset structure \cite{DeepSense}. Using the collected measurements/raw dataset, we built our development dataset for the radar-aided blockage prediction task. In this section, we describe our testbed, raw measurement database, and development dataset.

\subsection{DeepSense Tesbed-3}
We adopt Testbed 3 of the DeepSense 6G dataset \cite{DeepSense} for the data collection. Testbed 3 comprises two units: (i) Unit 1, a fixed receiver acting as a basestation and (ii) Unit 2, a static transmitter. Unit 1 includes a $60$ GHz uniform linear array (ULA) with $M_\mathrm{A}=16$ elements, and an FMCW radar board (TI AWR2243BOOST) equipped with 3 transmitter and 4 receiver antennas. Meanwhile, Unit 2 comprises an omni-directional static transmitter. The phased array of Unit 1 utilizes an over-sampled beamforming codebook of $64$ receive beams. In the radar, only a single transmit antenna along with the 4 receive antennas are activated. The radar (chirp) parameters are selected based on the short-range radar example of TI \cite{dham2017programming}, providing a maximum range of $45$m and velocity of $56$ km/s. The bandwidth of the utilized chirp frame covers $B=750$ MHz bandwidth with a chirp slope of $\mu=15$ MHZ/us over $L=128$ chirps/frame and $S=256$ samples/chirp. Next, we detail the collection scenario and development dataset. In \figref{fig:setup_im_a}, a picture of the testbed is presented.

\subsection{DeepSense Scenario 30}
To evaluate our solution, we construct Scenario $30$ of the DeepSense 6G dataset \cite{DeepSense}. In this scenario, a base station is placed on the sidewalk of a road, directed towards the transmitter, which is placed on the other side of the road. The transmission is blocked when the buses, cars, bicycles and pedestrians are passing through the LOS path. The received power via each beamforming vector and radar measurements are saved continuously to be processed later. In the construction of the dataset, the beam providing the most power and the corresponding power level are saved as the optimal beamforming vector and the maximum power level. For labeling the blockage status of the samples, first, a threshold level for the maximum receive power level is determined. The samples providing power level below this threshold are considered as blockages, which are further confirmed manually through the inspection of the RGB images that are captured from the camera of Unit-1. The sampling periodicity is determined as $9$ samples/s. The other details of the testbed and raw measurement database is summarized in \tabref{table:scenario30}. We also illustrate the scenario details with the pictures in \figref{fig:datasetup}.

\subsection{Development Dataset} \label{sec:dataset:dev}
To build the development dataset for the considered blockage prediction task, we reduce the number of measurements by keeping only the data points relevant to the blockage, and generate the data samples of sub-sequences for the blockage task. In the following, we refer to each measurement as the data point, and each set of measurements for blockage prediction task as a sample. For the development dataset, we first filtered the raw dataset to keep only $36$ data points before a blockage and $10$ data points after a blockage, including the first blockage instance. The filtered dataset comprises $14624$ data points. These data points consist of $307$ unique blockage sequences, which are later utilized to construct the input time-series samples and blockage/status labels of different lengths. For the training, validation and test samples, the sequences are split via $70/20/10\%$ ratio to provide unseen blockage sequences in the test and validation sets. Further, an observation of $T_o=8$ and a prediction window of $T_p=10$, we generate the sub-sequences of data. These sub-sequences are later utilized to be fed into the machine learning model and to generate the labels, $b$, from the individual blockage status, $\widetilde b$. To emphasize, $T_p=10$ is a soft selection and used for the generation of the samples (sub-sequences), i.e., $\bX[t]$ and $\widetilde b[t+t_p]$ $\forall t_p \in \{1, \ldots, T_p\}$. The final development dataset comprises $6965$ training, $1808$ validation and $907$ test samples. The final labels, $b[t]$ are later generated for each sample based on the selected $T_p$ value, as presented in the section \sref{sec:results}.

\section{Evaluation Results} \label{sec:results}
In this section, we evaluate the proposed blockage prediction solutions, namely the object tracking and the LSTM/deep learning  based solutions. In the object tracking method, we do not utilize the observation period $T_o$ since the tracking updates are continuously performed after the initial detection of an object. To clarify, there is no input sequence size, which can be interpreted as $T_o=\infty$. The current states of the objects, that are updated through the data points, are adopted in the prediction of future blockages, as described in \sref{sec:trackingsol}. In the deep learning solution, we adopt the prepared sub-sequences with $T_o=8$, as described in \sref{sec:dataset}. For different values of simulation parameters, a different deep learning model is trained. The training is carried out for up to $30$ epochs with the Adam algorithm \cite{kingma2014adam} using a learning rate of $10^{-3}$. An early stopping criterion of $5$ epochs is adopted to stop the training and save the model with minimum validation loss. As the sampling periodicity of the scenario is designed as $9$ data points per second, the frame duration is taken as $\tau_f \approx 110$ ms.  We also note that, in the following results, only the results related to the test set are illustrated.

\begin{figure}[t]
	\centering
	\includegraphics[width=1\columnwidth]{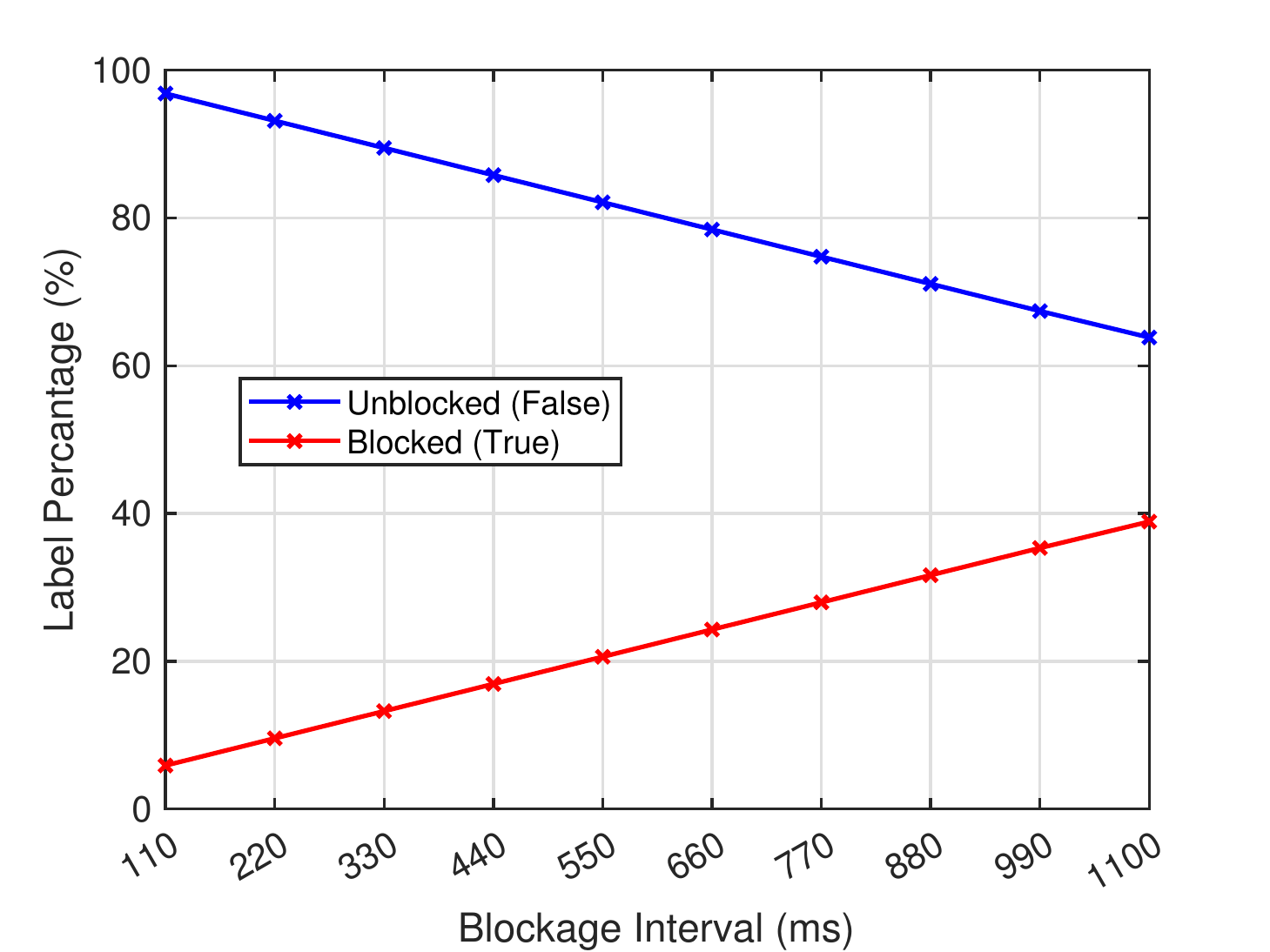}
	\caption{Distribution of the blockage status in the ground truth values and the predictions by the two proposed methods. The dataset is separated in a comparably balanced manner.}
	\label{fig:distribution}
\end{figure}

\textbf{Balance of the labels:} The balance of the labels is important for the machine learning tasks. It is usually preferable to have a balanced dataset. In the case it is not provided, different metrics and methods may be applied for a better performance and its evaluation. In the blockage prediction task, it is expected to have an imbalanced dataset with mostly unblocked samples. Although only the relevant samples are kept in our development dataset, the labels are generated based on the blockage interval, and the balance of the dataset changes based on this value. To evaluate the balance, we first investigate the balance for different blockage interval values ($T_p$). There are $951$ sub-sequences (samples) in test set. The percentage of the blocked (true) and unblocked (false) labels are illustrated in \figref{fig:distribution}. \textit{As seen in the figure, the dataset is imbalanced, especially for small blockage intervals, and a careful evaluation is required.} For this purpose, we adopt $F_1$ score as an evaluation metric. Specifically, $F_1$ score is defined as the harmonic mean of the precision and recall given by
\begin{equation}
	\text{Precision} = \frac{\text{TP}}{\text{TP}+\text{FP}},\quad  \text{Recall} = \frac{\text{TP}}{\text{TP}+\text{FN}}.
\end{equation}
where TP, FP and FN represent true positives, false positives and false negatives, respectively. The $F_1$ score provides a better metric for the evaluation of the imbalanced classification problems by penalizing the extreme values of precision and recall, which indicates the accuracy of true predictions and accuracy for predicting true labels. \textit{Further, in our simulations, we observe that the predictions from both the object detection and deep learning methods present a proportional set of prediction labels. This indicates a well-fit set of methods without requiring any extra attention.}

\begin{figure}[t]
	\centering
	\includegraphics[width=1\columnwidth]{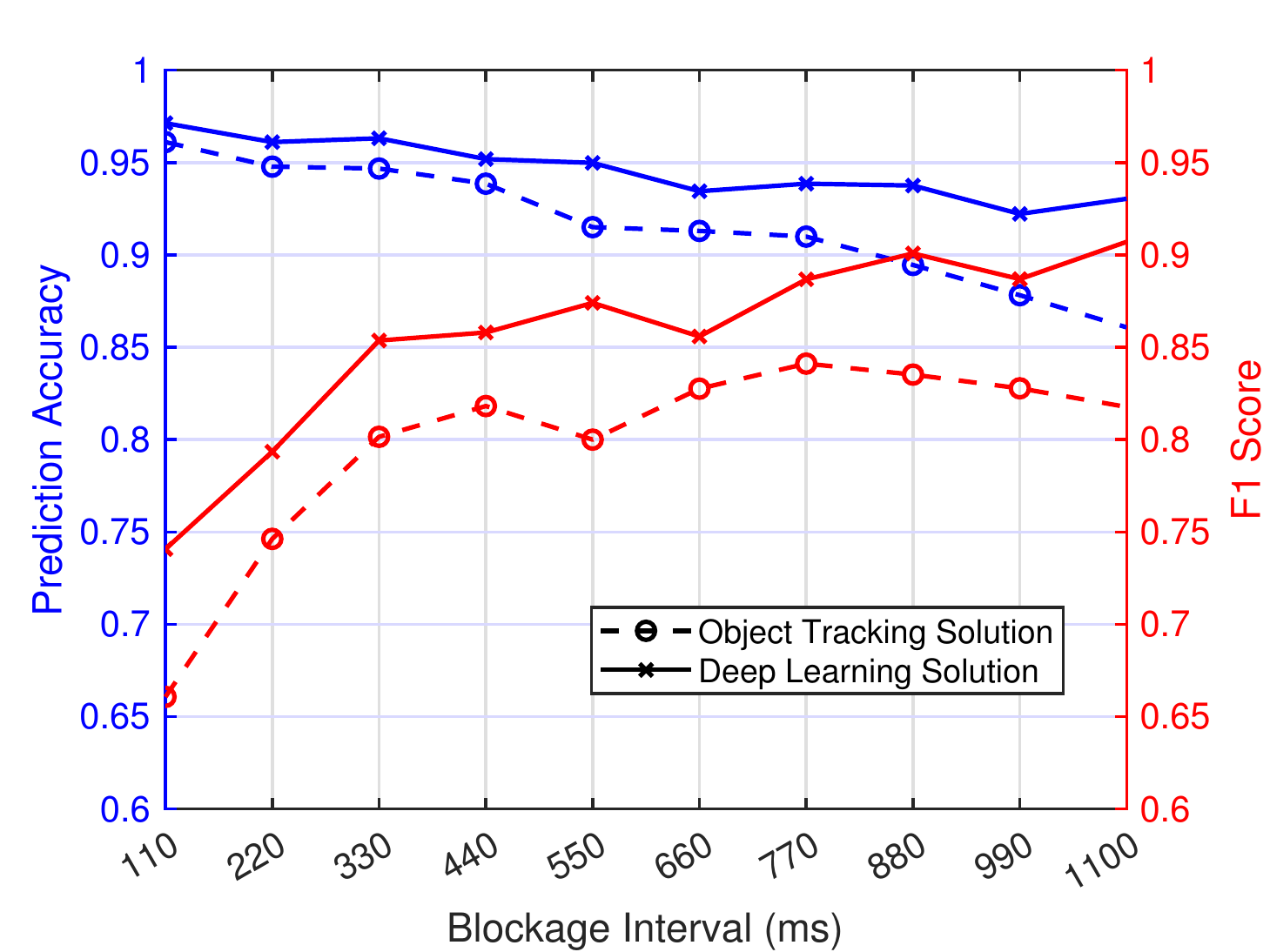}
	\caption{The performance of the proposed approaches for different blockage interval. Specifically, the frame duration is taken as $\tau_f=110$ ms and different number of blockage intervals, $T_p\in\{1, \dots, 10\}$ are shown.}
	\label{fig:accuracy}
\end{figure}

\textbf{Performance versus blockage interval:} In \figref{fig:accuracy}, we show the accuracy and $F_1$ score of the methodologies for different time blockage interval values ($T_p$). In the figure, the accuracy of the predictions present high-values with around $92-97\%$. However, the $F_1$ scores does not reflect similar results due the imbalance of the dataset. The $F_1$ score of the radar aided blockage prediction solutions increase with the larger blockage interval. Especially for $T_p \tau=110$ ms, both solutions perform poorly, potentially due to the low-angular resolution with $4$ radar receive antennas. It also shows more successful predictions into the future, while having more problems with the prediction of the closer blockages. In addition, the gap between the object tracking and deep learning solutions stays stable over different time intervals, potentially indicating a similar error difference between the solutions. \textit{The results highlights the delicate design requirements of the classical radar pipeline, while the deep learning solution easily attains significantly better results with the available data.}

\textbf{Complexity:} The direct inference duration comparison of the methods may not be fair due to the efficient implementation of the neural networks on the graphical processing units. However, we would like to comment on the complexity of the proposed solutions. In the object tracking method, the CFAR detection requires a 2D convolution (similar to a convolutional layer) with highly complex DBSCAN and Kalman filter applications of small number of variables. To clarify, the DBSCAN algorithm scales with the number of detected points and the Kalman filter scales with the number of tracked objects and states. \textit{In the deep learning solution, the neural network adopted in the evaluations only consists of $184,015$ parameters and presents a very small overhead to the system, especially with the specialized devices for the deep learning architectures.}

\section{Conclusion} \label{sec:conclusion}
In this paper, we proposed  radar aided blockage prediction approaches for mmWave and terahertz wireless networks. In particular, we developed two solutions based on (i) a standard radar object tracking pipeline along with a classifier based on the tracked object states, and (ii) a direct LSTM based deep learning solution using the range-angle maps. We evaluated our solutions based on a real-world large-scale dataset comprising co-existing radar and mmWave communication measurements. The results showed that the deep learning solution can achieve better blockage prediction results with a lower design complexity, while the object tracking solution comprises several layers of design and decision complexity.  For example, the results indicated that the proposed deep learning approach can predict future blockages 1 second before they happen with an $F1$ score of more than $90\%$ (and $> 90\%$ accuracy) while the object tracking method  presented $\sim 80\%$ accuracy. These results, among others,  demonstrate the promising gains of leveraging low-cost radar sensors to proactively predict blockages and enhance the reliability of mmWave networks. 

\balance


\begin{thebibliography}{10}
	\providecommand{\url}[1]{#1}
	\csname url@samestyle\endcsname
	\providecommand{\newblock}{\relax}
	\providecommand{\bibinfo}[2]{#2}
	\providecommand{\BIBentrySTDinterwordspacing}{\spaceskip=0pt\relax}
	\providecommand{\BIBentryALTinterwordstretchfactor}{4}
	\providecommand{\BIBentryALTinterwordspacing}{\spaceskip=\fontdimen2\font plus
		\BIBentryALTinterwordstretchfactor\fontdimen3\font minus
		\fontdimen4\font\relax}
	\providecommand{\BIBforeignlanguage}[2]{{%
			\expandafter\ifx\csname l@#1\endcsname\relax
			\typeout{** WARNING: IEEEtran.bst: No hyphenation pattern has been}%
			\typeout{** loaded for the language `#1'. Using the pattern for}%
			\typeout{** the default language instead.}%
			\else
			\language=\csname l@#1\endcsname
			\fi
			#2}}
	\providecommand{\BIBdecl}{\relax}
	\BIBdecl
	
	\bibitem{heath2016overview}
	R.~W. Heath, N.~Gonzalez-Prelcic, S.~Rangan, W.~Roh, and A.~M. Sayeed, ``An
	overview of signal processing techniques for millimeter wave {MIMO}
	systems,'' \emph{IEEE journal of selected topics in signal processing},
	vol.~10, no.~3, pp. 436--453, 2016.
	
	\bibitem{rappaport2019wireless}
	T.~S. Rappaport, Y.~Xing, O.~Kanhere, S.~Ju, A.~Madanayake, S.~Mandal,
	A.~Alkhateeb, and G.~C. Trichopoulos, ``Wireless communications and
	applications above 100 {GHz}: Opportunities and challenges for {6G} and
	beyond,'' \emph{IEEE access}, vol.~7, pp. 78\,729--78\,757, 2019.
	
	\bibitem{Alkhateeb_blockages}
	A.~Alkhateeb, I.~Beltagy, and S.~Alex, ``Machine learning for reliable mmwave
	systems: Blockage prediction and proactive handoff,'' in \emph{2018 IEEE
		Global Conference on Signal and Information Processing (GlobalSIP)}, 2018,
	pp. 1055--1059.
	
	\bibitem{alrabeiah2020deep}
	M.~Alrabeiah and A.~Alkhateeb, ``Deep learning for {mmWave} beam and blockage
	prediction using sub-6 {GHz} channels,'' \emph{IEEE Transactions on
		Communications}, vol.~68, no.~9, pp. 5504--5518, 2020.
	
	\bibitem{wu2021blockage}
	S.~Wu, M.~Alrabeiah, C.~Chakrabarti, and A.~Alkhateeb, ``Blockage prediction
	using wireless signatures: Deep learning enables real-world demonstration,''
	\emph{arXiv preprint arXiv:2111.08242}, 2021.
	
	\bibitem{charan2021visionjournal}
	G.~Charan, M.~Alrabeiah, and A.~Alkhateeb, ``Vision-aided {6G} wireless
	communications: Blockage prediction and proactive handoff,'' \emph{IEEE
		Transactions on Vehicular Technology}, pp. 1--1, 2021.
	
	\bibitem{charan2021visionposition}
	G.~Charan, T.~Osman, A.~Hredzak, N.~Thawdar, and A.~Alkhateeb,
	``Vision-position multi-modal beam prediction using real millimeter wave
	datasets,'' \emph{arXiv preprint arXiv:2111.07574}, 2021.
	
	\bibitem{wu2021blockageLiDAR}
	S.~Wu, C.~Chakrabarti, and A.~Alkhateeb, ``Lidar-aided mobile blockage
	prediction in real-world millimeter wave systems,'' \emph{arXiv preprint
		arXiv:2111.09581}, 2021.
	
	\bibitem{Taha2021}
	A.~Taha, Q.~Qu, S.~Alex, P.~Wang, W.~L. Abbott, and A.~Alkhateeb, ``Millimeter
	wave {MIMO}-based depth maps for wireless virtual and augmented reality,''
	\emph{IEEE Access}, vol.~9, pp. 48\,341--48\,363, 2021.
	
	\bibitem{Kumari_2018}
	P.~Kumari, J.~Choi, N.~González-Prelcic, and R.~W. Heath, ``{IEEE
		802.11ad}-based radar: An approach to joint vehicular communication-radar
	system,'' \emph{IEEE Transactions on Vehicular Technology}, vol.~67, no.~4,
	pp. 3012--3027, 2018.
	
	\bibitem{wan2000unscented}
	E.~A. Wan and R.~Van Der~Merwe, ``The unscented kalman filter for nonlinear
	estimation,'' in \emph{Proceedings of the IEEE 2000 Adaptive Systems for
		Signal Processing, Communications, and Control Symposium (Cat. No.
		00EX373)}.\hskip 1em plus 0.5em minus 0.4em\relax Ieee, 2000, pp. 153--158.
	
	\bibitem{DeepSense}
	\BIBentryALTinterwordspacing
	A.~Alkhateeb, G.~Charan, M.~Alrabeiah, T.~Osman, A.~Hredzak, N.~Srinivas, and
	M.~Seth, ``{DeepSense 6G}: A large-scale real-world multi-modal sensing and
	communication dataset,'' \emph{available on arXiv}, 2021. [Online].
	Available: \url{https://www.DeepSense6G.net}
	\BIBentrySTDinterwordspacing
	
	\bibitem{iovescu2017radarfundamentals}
	C.~Iovescu and S.~Rao, ``The fundamentals of millimeter wave sensors,''
	\emph{Texas Instruments}, pp. 1--8, 2017.
	
	\bibitem{dham2017programming}
	V.~Dham, ``Programming chirp parameters in {TI} radar devices,''
	\emph{Application Report SWRA553, Texas Instruments}, 2017.
	
	\bibitem{Zhang2021}
	Y.~Zhang, M.~Alrabeiah, and A.~Alkhateeb, ``Reinforcement learning of beam
	codebooks in millimeter wave and terahertz {MIMO} systems,'' \emph{IEEE
		Transactions on Communications}, pp. 1--1, 2021.
	
	\bibitem{rappaport2013millimeter}
	T.~S. Rappaport, S.~Sun, R.~Mayzus, H.~Zhao, Y.~Azar, K.~Wang, G.~N. Wong,
	J.~K. Schulz, M.~Samimi, and F.~Gutierrez, ``Millimeter wave mobile
	communications for {5G} cellular: It will work!'' \emph{IEEE access}, vol.~1,
	pp. 335--349, 2013.
	
	\bibitem{rohling1983radar}
	H.~Rohling, ``Radar cfar thresholding in clutter and multiple target
	situations,'' \emph{IEEE transactions on aerospace and electronic systems},
	no.~4, pp. 608--621, 1983.
	
	\bibitem{ester1996density}
	M.~Ester, H.-P. Kriegel, J.~Sander, X.~Xu \emph{et~al.}, ``A density-based
	algorithm for discovering clusters in large spatial databases with noise.''
	in \emph{kdd}, vol.~96, no.~34, 1996, pp. 226--231.
	
	\bibitem{schubert2017dbscan}
	E.~Schubert, J.~Sander, M.~Ester, H.~P. Kriegel, and X.~Xu, ``Dbscan revisited,
	revisited: why and how you should (still) use dbscan,'' \emph{ACM
		Transactions on Database Systems (TODS)}, vol.~42, no.~3, pp. 1--21, 2017.
	
	\bibitem{schubert2008comparison}
	R.~Schubert, E.~Richter, and G.~Wanielik, ``Comparison and evaluation of
	advanced motion models for vehicle tracking,'' in \emph{2008 11th
		international conference on information fusion}.\hskip 1em plus 0.5em minus
	0.4em\relax IEEE, 2008, pp. 1--6.
	
	\bibitem{vaishnav2020continuous}
	P.~Vaishnav and A.~Santra, ``Continuous human activity classification with
	unscented kalman filter tracking using {FMCW} radar,'' \emph{IEEE Sensors
		Letters}, vol.~4, no.~5, pp. 1--4, 2020.
	
	\bibitem{sainath2015convolutional}
	T.~N. Sainath, O.~Vinyals, A.~Senior, and H.~Sak, ``Convolutional, long
	short-term memory, fully connected deep neural networks,'' in \emph{2015 IEEE
		international conference on acoustics, speech and signal processing
		(ICASSP)}.\hskip 1em plus 0.5em minus 0.4em\relax IEEE, 2015, pp. 4580--4584.
	
	\bibitem{albawi2017understanding}
	S.~Albawi, T.~A. Mohammed, and S.~Al-Zawi, ``Understanding of a convolutional
	neural network,'' in \emph{2017 International Conference on Engineering and
		Technology (ICET)}.\hskip 1em plus 0.5em minus 0.4em\relax IEEE, 2017, pp.
	1--6.
	
	\bibitem{demirhan2021radarbeam}
	U.~Demirhan and A.~Alkhateeb, ``Radar aided {6G} beam prediction: Deep learning
	algorithms and real-world demonstration,'' \emph{arXiv preprint
		arXiv:2111.09676}, 2021.
	
	\bibitem{gers2000learning}
	F.~A. Gers, J.~Schmidhuber, and F.~Cummins, ``Learning to forget: Continual
	prediction with {LSTM},'' \emph{Neural computation}, vol.~12, no.~10, pp.
	2451--2471, 2000.
	
	\bibitem{akita2019object}
	T.~Akita and S.~Mita, ``Object tracking and classification using
	millimeter-wave radar based on {LSTM},'' in \emph{2019 IEEE Intelligent
		Transportation Systems Conference (ITSC)}.\hskip 1em plus 0.5em minus
	0.4em\relax IEEE, 2019, pp. 1110--1115.
	
	\bibitem{choi2019short}
	J.-W. Choi, S.-J. Ryu, and J.-H. Kim, ``Short-range radar based real-time hand
	gesture recognition using {LSTM} encoder,'' \emph{IEEE Access}, vol.~7, pp.
	33\,610--33\,618, 2019.
	
	\bibitem{kingma2014adam}
	D.~P. Kingma and J.~Ba, ``Adam: A method for stochastic optimization,''
	\emph{arXiv preprint arXiv:1412.6980}, 2014.
	
\end{thebibliography}
\end{document}